\documentclass[11pt,twoside]{article}

\usepackage{asp2004}
\usepackage{epsf}
\usepackage{psfig}
\usepackage{lscape}
\usepackage{graphicx}

\begin{document}
\title{A Dust Emission Model for Very Young Galaxies: \\
Expected Properties and Far Infrared Diagnostics}   
\author{Tsutomu T.\ Takeuchi}   
\affil{Laboratoire d'Astrophysique de Marseille, Traverse du Siphon BP8
13376, Marseille Cedex 12, France}    

\begin{abstract} 
Dust plays crucial roles in galaxy formation and evolution.
In the early epoch of galaxy evolution dust is only supplied by 
supernovae (SNe).
With the aid of a new physical model of dust production by SNe, 
we constructed a model of dust emission from forming galaxies.
We show the evolution of the spectral energy distribution (SED).
Then we adopt this model to a local starbursting dwarf galaxy
SBS~0335$-$052.
Further we discuss the SEDs of high redshift galaxies, and 
consider their observational feasibility.
\end{abstract}

\section{Introduction}

Active star formation (SF) is followed by heavy element 
production from the birth and death of stars.
Since the produced heavy elements generally exist in the form of 
dust grains, the dust grains absorb ultraviolet (UV) light
and re-emit it in the far infrared (FIR). 

Indeed, there is an extreme category of high-$z$ galaxies which have
large amount of dust and are extremely luminous in the FIR and submillimeter 
(submm) wavelengths.
Heavily hidden SF is suggested in these galaxies 
\citep[e.g.,][]{takeuchi01a,takeuchi01b}.
By examining the luminosity functions (LFs) at UV and FIR from 
{\sl GALEX} and {\sl IRAS}/{\sl Spitzer}, 
\citet{takeuchi05c} proved that the FIR LF shows much stronger evolution
than that of UV, though both evolve very strongly.
This indicates that the fraction of hidden SF rapidly increases toward
higher redshifts up to $z \sim 1$.
There is another important observable closely related to the dust 
emission from galaxies: the cosmic IR background (CIB).
Recently, \citet{takeuchi06} constructed the IR spectral
energy distribution (SED) of the Local Universe. 
The energy emitted in the IR is 25--30\% of the total energy 
budget.
In contrast, the IR (from near/mid-IR to millimeter) contribution 
is roughly (or even more than) a half in the CIB spectrum 
\citep[e.g.,][]{dole06}.
This also suggests a strong evolution of the IR contribution to 
the cosmic SED in the Universe.
Thus, understanding the radiative physics of dust is a fundamental 
task to have an unbiased view of the cosmic SF history.

In general, the IR observational data are obtained from photometric 
measurements.
Then, estimating the total IR luminosity from fluxes at discrete 
photometric bands is an essential step to discuss the energy budget
released in the UV and FIR.
Since \citet{takeuchi05a} demonstrated good linear relations between MIR 
and total IR luminosities for several orders of magnitudes in luminosity,
We may expect to have a good estimator, at least with a precision with 
a factor of five.
However, as well known, this is not a trivial task.
Various estimators have been proposed based on {\sl IRAS} bands, including
the classical FIR 
\citep[using 60 and $100\;\mu$m:][]{helou88},
Dale et al.'s TIR
\citep[60 and $100\;\mu$m:][]{dale01b},
and its revised version 
\citep[25, 60, and $100\;\mu$m:][]{dale02},
and Sanders \& Mirabel's IR
\citep[all {\sl IRAS} bands:][]{sanders96}.
\citet{takeuchi05a} examined these four estimators using a galaxy sample 
with known SEDs, and showed that they work well for normal galaxies.
However, there are a few categories of galaxies for which the FIR-based
ones (the former two) do not give a good estimate. 
We focus on one of these galaxies, i.e., actively star-forming galaxies,
and try to model their SEDs.

Throughout this paper, we use a cosmological parameter set of 
$(h,\Omega_0,\lambda_0)=(0.7,0.3,0.7)$, where 
$h\equiv H_0/100 \;[\mbox{km\,s}^{-1}\mbox{Mpc}^{-1}]$.

\section{SED Model for Forming Galaxies}\label{sec:model}

\subsection{Species and size distribution of dust grains produced by SNe II 
}\label{subsec:dust_species}

\citet[][hereafter N03]{nozawa03} investigated the formation of dust 
grains in the ejecta of Population III SNe 
(SNe II and PISNe, whose progenitors are initially metal-free), 
taking into account the following aspects:
(i) the time evolution of gas temperature is calculated by solving 
the radiative transfer equation including the energy deposition of 
radioactive elements.
(ii) the radial density profile of various metals is properly
considered, and 
(iii) unmixed and uniformly mixed cases in the He core are considered.
In the unmixed case, the original onion-like structure of elements is 
preserved, and in the mixed case, all the elements are uniformly mixed
in the helium core.
\citet{takeuchi05c} showed that the unmixed dust production is 
preferred to reproduce the SED of a local starbursting dwarf galaxy
SBS~0335$-$052 (discussed below).
In addition, \citet{hirashita05} proved that the unmixed scenario 
can also reproduce the extinction curve of a high-redshift quasar.
Hence, in the following, we only discuss the unmixed case.
The size of the grains spans a range of three orders of magnitude, 
depending on the grain species. 
The size spectrum summed up over all the grain species has a very 
broad distribution, and very roughly speaking, it might be approximated by
a power law.

\subsection{Star formation, chemical evolution, and dust production}

For constructing the chemical evolution model of a young galaxy, 
we adopt the following assumptions:
\begin{enumerate}
\item We use a closed-box model, i.e., we neglect an infall and outflow of gas 
in the scale of a star-forming region.
\item For the initial mass function (IMF), we adopt the Salpeter IMF 
\citep{salpeter55}: $\phi(m) \propto m^{-2.35}$, 
with mass range of $(m_{\rm l}, m_{\rm u})= (0.1\;M_\odot,100\;M_\odot)$.
\item We neglect the contribution of SNe Ia and winds from low-mass 
evolved stars to the formation of dust, because we consider the timescale 
younger than $10^9\;\mbox{yr}$.
\item The interstellar medium is treated as one zone, and the growth of 
dust grains by accretion is neglected.
Within the short timescale considered here, it can be assumed safely.
\item We also neglect the destruction of dust grains within the young age
considered \citep[see e.g.,][]{jones96}.
\item We assumed a constant SFR for simplicity.
\end{enumerate}
Using these assumptions, we calculate the chemical evolution.

\subsection{SED construction}\label{subsec:sed_construction}

In this subsection, we present the construction of the SED from dust.
All the details of the calculations are presented in 
\citet[][]{takeuchi03a},
\citet[][]{takeuchi04b}, and
\citet[][]{takeuchi05a}.
(hereafter T03, T04 and T05, respectively).
\begin{enumerate}
\item{\sl Stochastic heating of very small grains}

Very small grains cannot establish thermal equilibrium with the ambient 
radiation field.
This is called stochastic heating. 
{}To treat the effect, we applied 
a multidimensional Debye model \citep[e.g.,][]{draine01} to the 
specific heats of the grain species.

\item{\sl Emission}

The emission from dust is calculated basically according to \citet{draine85}
(see T03).
Total dust emission is obtained as a superposition of the emission from 
each grain species.
We constructed $Q(a, \lambda)$ of each grain species from available 
experimental data via Mie theory.

\item{\sl Extinction}

Self-absorption in the MIR for a very optically thick case is treated 
by a thin shell approximation (see T04).
\end{enumerate}

\section{Results}\label{sec:results}

\begin{figure}[!ht]
\centering\includegraphics[angle=90,width=0.9\textwidth]{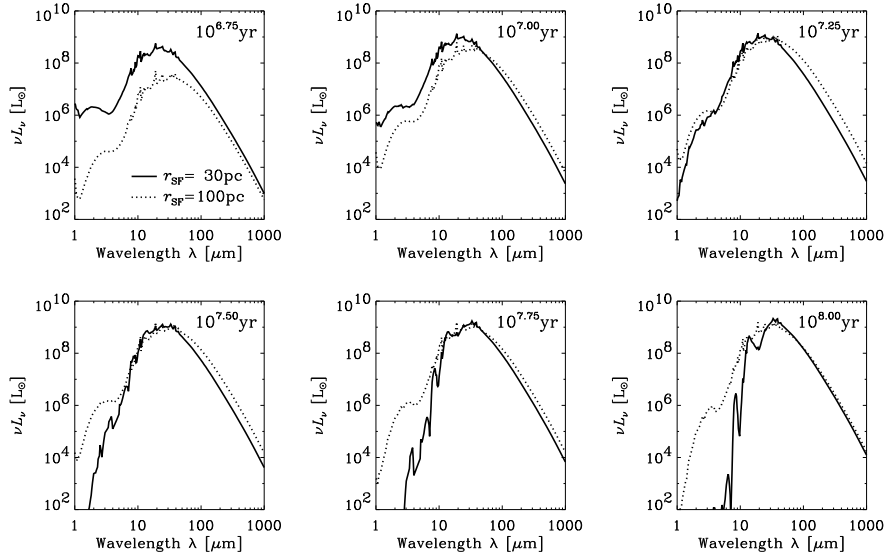}
\caption{
The evolution of the infrared (IR) spectral energy distribution 
(SED) of a very young galaxy.
The sizes of the star forming region, $r_{\rm SF}$, is 30~pc (solid lines) 
and 100~pc (dotted lines).
}\label{fig:sed}
\end{figure}

\subsection{Evolution of infrared SED}

We first show the evolution of the IR SED of forming galaxies.
For these calculation we adopted the star formation rate 
${\rm SFR}=1\,M_\odot\,\mbox{yr}^{-1}$.
We adopt $r_{\rm SF} = 30$~pc and 100~pc.
These values are relevant when describing `dwarf-like' young galaxies. 
The results are presented in Figure~\ref{fig:sed}\footnote{
The SED data are available at
{\tt http://www.kwasan.kyoto-u.ac.jp/$\sim$takeuchi/dust.html}.}.
We calculated the evolution of the SED in the age range of 
$10^{6.5}\mbox{--}10^{8}$~yr.
After $10^{7.25}$~yr, the N--MIR continuum is extinguished by the 
self-absorption in the case of $r_{\rm SF}=30$~pc. 
In contrast, the self-absorption is not significant for $r_{\rm SF}=100$~pc.
The SEDs have their peaks at a wavelength $\lambda \simeq 
20\mbox{--}30\;\mu$m, which is much shorter than those of dusty giant galaxies
at $z=1\mbox{--}3$ detected by SCUBA.

\section{Discussion}\label{sec:discussion}

\subsection{Nearby forming dwarf galaxy SBS~0335$-$052}\label{subsec:local}

It is still a difficult task to observe galaxies in their very first phase 
of the SF, especially to detect their dust emission directly.
Since a recent observation of SBS~0335$-$052\ by {\sl Spitzer} has
been reported \citep{houck04}, it is timely to reconsider these 
`textbook objects' with new datasets.
In addition, understanding the SEDs of these objects will shed light to 
the physics of interstellar matter and radiation of high-$z$ galaxies also 
via empirical studies \citep[e.g.,][]{takeuchi03b,takeuchi05b}.

SBS~0335$-$052\ is a local galaxy ($\sim 54\;\mbox{Mpc}$) with 
$\mbox{SFR} = 1.7 \;M_\odot\,\mbox{yr}^{-1}$ \citep{hunt01} and
extremely low metallicity $Z = 1/41\,Z_\odot$.
This galaxy is known to have an unusual IR SED and strong flux at N--MIR.
It has a very young starburst ($\mbox{age} \la 5\,\mbox{Myr}$) without 
significant underlying old stellar population \citep{vanzi00}.
T03 have modeled the SED of SBS~0335$-$052\ and reported a good
agreement with the available observations at that time.
However, \citet{houck04} presented new data of the MIR SED by {\sl Spitzer},
and reported a deviation of the model by a factor of two or three.
Their observation indicated that SBS~0335$-$052\ has even 
more FIR-deficient SED than ever thought.
Hence, it is interesting to examine whether our present model can reproduce
the extreme SED of this galaxy.

\begin{figure}[!ht]
\centering\includegraphics[width=0.45\textwidth]{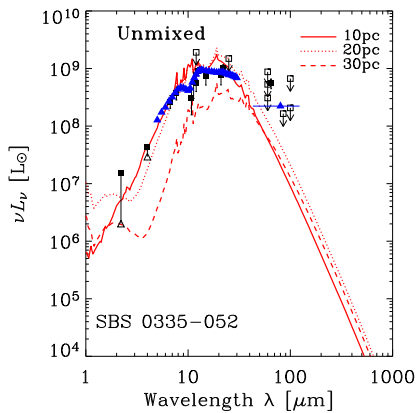}
\caption{The model for the SED of a nearby star-forming dwarf galaxy SBS~0335$-$052.
}\label{fig:sed_sbs}
\end{figure}

We show the model SEDs for SBS~0335$-$052\ in Figure~\ref{fig:sed_sbs}.
We have calculated the SED for $r_{\rm SF}=10$, 20, and 30~pc.
The SFR is fixed to be $1.7\,M_\odot\,\mbox{yr}^{-1}$, and the age is 
$10^{6.5}$~yr.
Details of the observational data are found in T03 and T05.
In the FIR regime, our model SEDs are consistent with the strong constraint 
given by \citet{houck04}.
At MIR, though we cannot give an excellent fit to the observed data, the model
SEDs roughly agree with them.
The very strong N--MIR continuum of SBS~0335$-$052\ is well reproduced.
The dust mass calculated by our model at this age of SBS~0335$-$052\ is 
$1\mbox{--}2 \times 10^3\;M_\odot$, consistent with the observationally
estimated value by \citet{dale01a}.
We note that the mass estimation is strongly dependent on the assumed
dust species and their emissivities, and grain size distribution.
As we discussed in Section~\ref{sec:results}, the continuum radiation in the
N--MIR is dominated by stochastically heated dust emission, which is
completely different from modified blackbody.
Therefore, when we try to estimate the dust mass, we must take care to 
determine the corresponding grain properties, i.e., radiative processes
and grain which are related to the observed SED of galaxies.

\subsection{Lyman-break galaxies}

Even in LBGs, there is clear evidence that they contain 
non-negligible amount of dust \citep[e.g.,][]{adelberger00}. 
A high dust temperature ($\ga 70\,$K) is suggested by subsequent studies
\citep[e.g.,][]{ouchi99,chapman00,sawicki01}.
{}To make a consistent picture of the dust emission from LBGs, 
we investigate the expected appearance of the LBGs with 
an improved dust grain formation of N03.

We set the input parameters of the SED model for LBGs as follows.
The SFR of LBGs spreads over the range of 
$\mbox{SFR} \simeq 1 \mbox{--} 300\,M_\odot \mbox{yr}^{-1}$ with a median 
of $\mbox{SFR} \simeq 20\,M_\odot \mbox{yr}^{-1}$ 
\citep[e.g., ][]{erb03}.
Thus, the basic framework of the T03 model is also valid for LBGs.
In this work, we consider the moderate case of 
$\mbox{SFR}=30\,M_\odot \mbox{yr}^{-1}$ over the age of $10^{6.5}
\mbox{--}10^{8}$~yr.
The most important information to calculate the IR SED is the effective
size of the star forming region, but it is the most uncertain quantity
(see T04).
Since the mean half-light radius of LBGs is estimated to be $\sim 1.6\,
\mbox{kpc}$ from {\sl HST} observations \citep{erb03}, 
we use the galaxy radius as the radius of a star-forming region, 
and set $r_{\rm SF} = 2\,\mbox{kpc}$ according to T04.

\begin{figure}[!ht]
\centering\includegraphics[angle=90,width=0.9\textwidth]{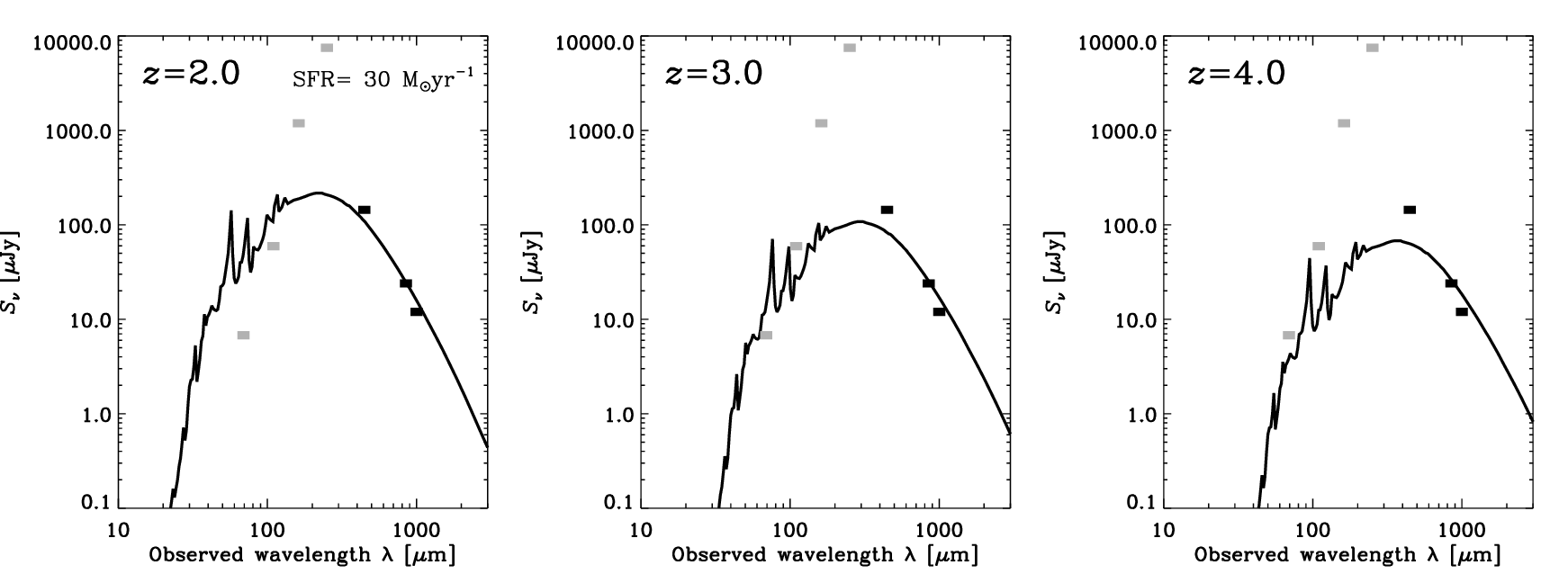}
\caption{The expected flux densities of an LBG.
In this figure we set the galaxy burst age of $t=10^8\,\mbox{yr}$.
}\label{fig:lbg_obs}
\end{figure}
We show the observed IR/submm SEDs of LBGs at $z=2$, 3, and 4 in 
Figure~\ref{fig:lbg_obs}.
For simplicity we only show the SED with the age of $10^8$~yr.
The thick black short horizontal lines indicate the 3-$\sigma$ detection limits
for 8-hour observation by ALMA (Atacama Large Millimeter Array).
Here we assumed 64 antennas and three wavelength bands, 450, 850, and 
$1080\,\mu$m.
We also show the 3-$\sigma$ source confusion limit of {\sl Herschel}
at 75, 160, 250, and 350$\,\mu$m bands by thick gray horizontal lines.
These limits are based on `the photometric criterion' of \citep{lagache03}
\citep[see also ][]{takeuchi04a}.

As discussed in T04, the detectability of LBGs is not strongly dependent on
their redshifts.
Detection at $350\;\mu$m seems impossible for moderate-SFR LBGs.
However at longer wavelengths, 
if the age $\ga 10^8\,{\rm yr}$ and $\mbox{SFR} \ga 30\,M_\odot\,
{\rm yr}^{-1}$, LBGs can be detected at a wide range of redshifts in the submm
by ALMA deep survey.
In the FIR, {\sl Herschel} will detect the dust emission from LBGs at 
$z\simeq 2$, but difficult at higher-$z$.

\subsection{Toward higher redshifts}

\begin{figure}[!ht]
\centering\includegraphics[angle=90,width=0.9\textwidth]{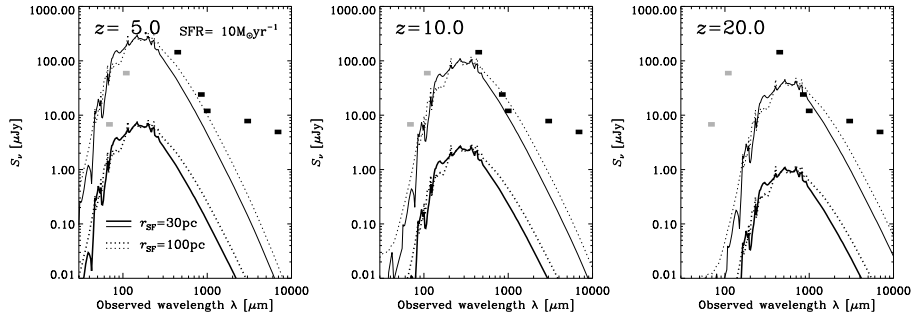}
\caption{The expected flux densities of a dwarf star-forming galaxy located at
$z =5, 10$, and 20. 
Solid lines represent the SEDs for $r_{\rm SF}=30\;\mbox{pc}$
and dotted lines for $r_{\rm SF}=100\;\mbox{pc}$.
}\label{fig:dwarf_highz}
\end{figure}

Based on the above discussions, we give a brief consideration on 
the observation of very high-$z$ galaxies ($z \ga 5$) here.
In modern hierarchical structure formation scenarios, it would be more 
reasonable to assume a small, subgalactic clump as a first forming galaxy.
Consider a dark halo of mass $\sim 10^9\;M_\odot$, then it is expected to 
contain a gas with mass $\simeq 10^8\;M_\odot$.
If gas collapses on the free-fall timescale with an efficiency of 
$\epsilon_{\rm SF}$ (we assume $\epsilon_{\rm SF}=0.1$), we obtain
the following evaluation of the SFR \citep{hirashita04}:
\begin{eqnarray}
  \mbox{SFR} \simeq 0.1 \left(\frac{\epsilon_{\rm SF}}{0.1}\right)
    \left(\frac{M_{\rm gas}}{10^7\;M_\odot}\right)^{3/2}
    \left(\frac{r_{\rm SF}}{100\;\mbox{pc}}\right)^{-3/2} \;
    [M_\odot \mbox{yr}^{-1}]\;.
\end{eqnarray}
If $M_{\rm gas}\simeq 10^8\;M_\odot$, we have $\mbox{SFR}\simeq
3(r_{\rm SF}/100\;\mbox{pc})^{-3/2}\;M_\odot\,\mbox{yr}^{-1}$.
In addition, an extremely high-$z$ galaxy observed by 
{\sl HST} has a very compact morphology \citep{kneib04}.
We also mention that, from a theoretical side, high-$z$ galaxies are 
suggested to be dense and compact compared to nearby galaxies 
\citep[e.g.,][]{hirashita02b}.
Thus, we consider a dwarf galaxy with $\mbox{SFR} = 10\;M_\odot\mbox{yr}^{-1}$
as an example, and we adopt $r_{\rm SF} = 30$~pc and 100~pc.
The age is set to be $10^7$~yr.
If the age is older, they will become easier to detect if a constant 
SFR takes place.
We show the expected SEDs for such galaxies at $z=5$, 10, and 20 in 
Figure~\ref{fig:dwarf_highz}.
As expected, it seems almost impossible to detect such objects 
by {\sl Herschel} or ALMA.

There is, however, a hope to observe such a small forming galactic clump
directly: gravitational lensing works very well as a natural huge 
telescope.
If we assume a lens magnification factor of 40, such a small galaxy becomes
detectable (thin lines in Figure~\ref{fig:dwarf_highz}).
Since the expected SED of such a compact dwarf galaxy has a strong MIR 
continuum at their rest frame, it can be feasible to detect at the FIR in 
the observed frame.
A cooled FIR space telescope is more suitable for such observations, 
and this will be a strong scientific motivation for a future project 
like {\sl SPICA}.
At higher-$z$, they can be detectable by ALMA survey.
Even at $z \simeq 20$, they can be detected by a standard 8-hour survey
of ALMA, if a lensing takes place.
{}For a practical use, we must estimate how frequently such lensing events 
occur for high-$z$ objects.
Suppose a cluster of galaxies at $z_{\rm l}\simeq 0.1\mbox{--}0.2$ 
whose dynamical mass $M_{\rm dyn}$ is $5 \times 10^{14}\,M_\odot$ 
and whose mass distribution obeys the singular isothermal sphere.
We denote the strong lensing cross section, i.e., the area of the region 
in the source plane for which the resulting magnification by a cluster 
is larger than $\mu$, as $\sigma(>\mu)$.
\citet[][]{perrotta02} presented $\sigma(>\mu)$ as a function of $M_{\rm dyn}$
for $z_{\rm lens}=1.0$.
Since $\sigma(>\mu)\propto {D_{\rm ls}}^2$ ($D_{\rm ls}$ is the 
angular-diameter distance between the lens and the source),
we can convert their result to our condition and obtain 
$\sigma(>10) \simeq 30\;\mbox{arcsec}^2$ on the source plane.
This result is almost independent of the source redshifts.
Setting the limiting flux density $S_\nu = 1\;\mu$Jy and using the number 
counts of \citet{hirashita02b} for galaxies at $z>5$, 
we have an expected number of galaxies 
suffering a strong lensing to be $\simeq 1\mbox{--}3$.
Thus, we expect at least a few strongly lensed IR galaxies to this survey
depth.

\section{Conclusion}\label{sec:conclusion}

With the aid of a new physical model of dust production by SNe developed 
by N03, we constructed a model of dust emission from a very young galaxies.

The SED of a local starbursting dwarf galaxy, SBS~0335$-$052, was calculated.
Our present model SED naturally reproduces the strong N--MIR continuum 
and the lack of cold FIR emission of SBS~0335$-$052.
Then we calculated the evolution of the SED of LBGs.
Finally, we considered the observations of forming galaxies at $z \ga 5$.
For small forming galaxies with a gas mass $M_{\rm gas}\simeq 10^{8}\;
M_\odot$, it is almost impossible to detect their intrinsic flux by ALMA 
or {\sl Herschel}.
However, the gravitational lensing is found to be a very effective 
tool to detect such small star-forming galaxies at $z \ga 5$.

\acknowledgements 
This work is produced through fruitful collaborations with 
Takako T.\ Ishii, Hiroyuki Hirashita, Takashi Kozasa, 
Takaya Nozawa, Andrea Ferrara, Leslie K.\ Hunt, V\'eronique Buat, 
Denis Burgarella, Alessandro Boselli,
Jorge Iglesias-P\'aramo, Herv\'e Dole, Guilaine Lagache, 
Michel Dennefeld, and Jean-Loup Puget, among many others.
We thank all of them.


\begin{thebibliography}{}
\bibitem[Adelberger \& Steidel(2000)]{adelberger00}
 Adelberger, K.\ L., \& Steidel, C.\ C.\ 2000, ApJ, 544, 218

\bibitem[Chapman et al.(2000)]{chapman00}
 Chapman, S.\ C., et al.\ 2000, MNRAS, 319, 318

\bibitem[Dale et al.(2001a)]{dale01a}
 Dale, D.~A., Helou, G., Neugebauer, G., et al.\ 2001, \aj, 122, 1736 
 
\bibitem[Dale et al.(2001b)]{dale01b}
 Dale, D.~A., Helou, G., Contursi, A., et al.\ 2001, \apj, 549, 215 
 
\bibitem[Dale \& Helou(2002)]{dale02}
 Dale, D.~A., \& Helou, G.\ 2002, \apj, 576, 159 
 
\bibitem[Dole et al.(2006)]{dole06}
 Dole, H., Lagache, G., Puget, J.-L., et al., 2006, A\&A, in press
 (astro-ph/0603208)

\bibitem[Draine \& Lee(1984)]{draine84}
 Draine, B.\ T., \& Lee, H.\ M.\ 1984, ApJ, 285, 89

\bibitem[Draine \& Anderson(1985)]{draine85}
 Draine, B.\ T., \& Anderson, L.\ 1985, ApJ, 292, 494

\bibitem[Draine \& Li(2001)]{draine01}
 Draine, B.\ T., \& Li A.\ 2001, ApJ, 551, 807

\bibitem[Erb et al.(2003)]{erb03}
 Erb, D.\ K., et al.\ 2003, ApJ, 591, 101

\bibitem[Helou et al.(1988)]{helou88}
 Helou, G., Khan, I.~R., Malek, L., \& Boehmer, L.\ 1988, \apjs, 68, 151 
 
\bibitem[Hirashita \& Ferrara(2002)]{hirashita02b} 
 Hirashita, H., \& Ferrara, A., 2002, MNRAS, 337, 921

\bibitem[Hirashita \& Hunt(2004)]{hirashita04} 
 Hirashita, H., \& Hunt, L.\ K., 2004, A\&A, 421, 555

\bibitem[Hirashita et al.(2005)]{hirashita05} 
 Hirashita, H., Nozawa, T., Kozasa, T., et al.\ 2005, MNRAS, 357, 1077

\bibitem[Houck et al.(2004)]{houck04}
 Houck, J.\ R., et al.\ 2004, ApJS, 154, 211

\bibitem[Hunt, Vanzi, \& Thuan(2001)]{hunt01}
 Hunt, L.\ K., Vanzi, L., \& Thuan, T.\ X.\ 2001, ApJ, 377, 66

\bibitem[Jones, Tielens, \& Hollenbach(1996)]{jones96}
 Jones, A.\ P., Tielens, A.\ G.\ G.\ M., \& Hollenbach, D.\ J.\ 1996,
 ApJ, 469, 740

\bibitem[Kneib et al.(2004)]{kneib04} 
 Kneib, J., Ellis, R.~S., Santos, M.~R., Richard, J.\ 2004, ApJ, 607, 697 

\bibitem[Lagache, Dole, \& Puget(2003)]{lagache03}
 Lagache, G., Dole, H., Puget, J.-L., 2003, MNRAS, 338, 555

\bibitem[Nozawa et al.(2003)]{nozawa03}
 Nozawa, T., Kozasa T., Umeda H., et al.\ 2003, ApJ, 598, 785 (N03)

\bibitem[Ouchi et al.(1999)]{ouchi99}
 Ouchi, M., Yamada, T., Kawai, H., Ohta, K.\ 1999, ApJ, 517, L19

\bibitem[Perrotta et al.(2002)]{perrotta02}
 Perrotta, F., Baccigalupi, C., Bartelmann, M., et al.\ 2002, MNRAS, 329, 445 

\bibitem[Salpeter(1955)]{salpeter55}
 Salpeter, E.\ 1955, ApJ, 121, 161

\bibitem[Sanders \& Mirabel(1996)]{sanders96}
 Sanders, D.~B., \& Mirabel, I.~F.\ 1996, \araa, 34, 749 
 
\bibitem[Sawicki(2001)]{sawicki01}
 Sawicki, M.\ 2001, AJ, 121, 2405


\bibitem[Takeuchi et al.(2001a)]{takeuchi01a}
 Takeuchi, T.\ T., Ishii, T.\ T., Hirashita, H., et al.\ 2001a, PASJ, 53, 37

\bibitem[Takeuchi et al.(2001b)]{takeuchi01b}
 Takeuchi, T.\ T., Kawabe, R., Kohno, K., et al.\ 2001b, PASP, 113, 586

\bibitem[Takeuchi et al.(2003)]{takeuchi03a}
 Takeuchi, T.\ T., Ishii, T.\ T., Hirashita, H., et al.\ 2003, MNRAS,
 343, 839 (T03)

\bibitem[Takeuchi, Yoshikawa, \& Ishii(2003)]{takeuchi03b}
 Takeuchi, T.\ T., Yoshikawa, K., Ishii, T.\ T.\ 2003, ApJ, 587, L89

\bibitem[Takeuchi \& Ishii(2004a)]{takeuchi04a}
 Takeuchi, T.\ T., Ishii, T.\ T., 2004a, ApJ, 604, 40

\bibitem[Takeuchi \& Ishii(2004b)]{takeuchi04b}
 Takeuchi, T.\ T., Ishii, T.\ T., 2004b, A\&A, 426, 425 (T04)

\bibitem[Takeuchi et al.(2005a)]{takeuchi05a}
 Takeuchi, T.\ T., Buat, V., Iglesias-P\'aramo, J., et al.\ 2005a, A\&A,
 432, 423

\bibitem[Takeuchi et al.(2005b))]{takeuchi05b}
 Takeuchi, T.\ T., Ishii, T.\ T., Nozawa, T., et al.\ 2005b, MNRAS, 362,
 592 (T05)

\bibitem[Takeuchi, Buat, \& Burgarella(2005))]{takeuchi05c}
 Takeuchi, T.\ T., Buat, V., \& Burgarella, D.\ 2005, A\&A, 440, L17

\bibitem[Takeuchi et al.(2006)]{takeuchi06}
 Takeuchi, T.\ T., Ishii, T.\ T., Dole, H., et al.\ 2006, A\&A, 448, 525

\bibitem[Vanzi et al.(2000)]{vanzi00}
 Vanzi, L., Hunt, L.\ K., Thuan, T.\ X., Izotov, Y.\ I.\ 2000, A\&A, 363, 493
\end{thebibliography}
\end{document}